\PassOptionsToPackage{pdftex}{graphicx}
\PassOptionsToPackage{pdftex}{color}

\documentclass{jjap3}

\usepackage{txfonts}

\usepackage[sort&compress,numbers]{natbib}
\usepackage{doi}
\makeatletter
\providecommand{\ext@figure}{lof}
\providecommand{\ext@table}{lot}
\makeatother
\usepackage{hyperref}
\hypersetup{hidelinks} % 投稿用設定

\title{Noise Suppression for Time Difference of Arrival: Performance Evaluation of a Generalized Cross-Correlation Method Using Mean Signal and Inverse Filter}
\author{Hirotaka Obo$^{1}$\thanks{E-mail: obo.hirotaka468@naro.go.jp}, Yuki Fujita$^{1}$, Masahisa Ishii$^{1}$, Hideki Moriyama$^{1}$, Ryota Tsuchiya$^{1}$, Yuta Ohashi$^{1}$ and Kotaro Seki$^{1}$}
\inst{$^{1}$Institute for Rural Engineering, National Agriculture and Food Research Organization (NARO), Tsukuba, Ibaraki 305-8609, Japan}

\abst{This paper proposes a novel generalized cross-correlation (GCC) method, termed GCC-MSIF, to improve time difference of arrival (TDOA) estimation accuracy in noisy environments. Conventional GCC methods often suffer from performance degradation under low signal-to-noise ratio (SNR) conditions, particularly when the signal bandwidth is unknown. GCC-MSIF introduces a "mean signal" estimated from multi-channel inputs and an "inverse filter" to virtually reconstruct the source signal, enabling adaptive suppression of out-of-band noise. Numerical simulations simulating a small-scale array demonstrate that GCC-MSIF significantly outperforms conventional methods, such as GCC-PHAT and GCC-SCOT, in low SNR regions and achieves robustness comparable to or exceeding the maximum likelihood (GCC-ML) method. Furthermore, the estimation accuracy improves scalably with the number of array elements. These results suggest that GCC-MSIF is a promising solution for robust passive localization in practical blind environments.}

\begin{document}
\maketitle

\section{Introduction}

Time difference of arrival (TDOA) estimation \cite{Zekavat-2011-HandbookPositionLocation} is a fundamental technology for localization \cite{Obo-2022-2Pa24NarrowPitchMatrixType} and direction of arrival (DOA) estimation \cite{Wang-2013-HighResolutionDirectionArrival,Zempo-2012-DirectionArrivalEstimationBased,Zempo-2013-LocalizationAcousticReflectiveBoundary}. These technologies estimate the position or arrival direction of a signal source by precisely measuring the arrival time differences of signals received by multiple spatially separated sensors, such as microphone arrays \cite{Brandstein-2001-MicrophoneArraysSignalProcessing}. Key applications include sound source localization \cite{Iwase-2015-FinWhaleVocalizationsObserved,Grondin-2016-NoiseMaskTDOASound}. Sound source localization is considered an indispensable technology in a wide range of fields, including robot audition \cite{Ui-HyunKim-2008-SpeakerLocalizationUsingTDOAbased,Valin-2007-RobustLocalizationTrackingSimultaneous}, smart speakers \cite{Li-2017-AcousticModelingGoogleHome,Li-2021-SuperSoundcompassHighaccuracyAcousticLocalization}, and conferencing systems \cite{Tashev-2009-SoundCaptureProcessingPractical,Li-2021-SuperSoundcompassHighaccuracyAcousticLocalization}, where there is a growing demand for smaller and more precise systems \cite{Fischer-2024-EvaluationSparseAcousticArray,DaSilva-2017-DesignConsiderationsWhenAccelerating,Zhang-2014-AcousticSourceLocalizationSubspace}.

However, since TDOA estimation is used in situations where the transmitted signal is unknown, it faces the challenge of performance degradation due to the influence of noise superimposed on the received signals. While time of arrival (TOA) estimation \cite{Pahlavan-2002-IndoorGeolocationScienceTechnology,Gustafsson-2005-MobilePositioningUsingWireless,Thong-un-2015-ImprovementAirbornePositionMeasurementsa,Iwaya-2017-AcousticalPositioningMethodUsing} is applicable when the waveform of the transmitted signal is known, TDOA estimation is used in passive localization in many real-world environments because the signal is unknown. Consequently, in TDOA estimation, noise degrades the correlation between signals and causes errors in time difference estimation; thus, ensuring accuracy in noisy environments is a critical technical issue.

As a representative method to improve noise robustness in TDOA estimation, the generalized cross-correlation (GCC) method \cite{Knapp-1976-GeneralizedCorrelationMethodEstimation,Sim-2016-UnderwaterAcousticSourceLocalization}, which applies frequency weighting to the cross-spectrum, has been proposed. For example, the GCC-PHAT (phase transform) method, which utilizes only phase information, is known to exhibit good performance in reverberant environments. Additionally, the GCC-ML (maximum likelihood) method is known as a weighting function that provides the theoretically optimal estimate in Gaussian noise environments. However, the performance of both methods tends to deteriorate in environments with poor signal-to-noise ratios (SNR), creating a need for the construction of superior methods.

Therefore, in this paper, we propose a novel GCC method, termed GCC-MSIF, which introduces a mean signal and an inverse filter to improve noise suppression performance, particularly in low SNR environments. This method is characterized by estimating an average signal spectrum from signals received by multiple sensors and designing a filter based on this spectrum to effectively suppress noise components. Through this approach, we aim to estimate TDOA with high precision even in low SNR environments where estimation is difficult with conventional methods.

This paper is an extension of our previous work \cite{Obo-2025-1P211NoiseSuppressionMethod}, comprehensively demonstrating the effectiveness of GCC-MSIF by adding exhaustive comparisons with conventional methods and theoretical limits (CRLB). While the previous study presented the basic principle of the method and its effectiveness under limited conditions, this paper conducts a more systematic and extensive evaluation. The main contribution of this paper is to quantitatively clarify the advantages of GCC-MSIF compared to conventional technologies and the extent to which its performance approaches the theoretical limits.

The structure of this paper is as follows. Section 2 describes the theoretical background of GCC methods and details the derivation process of GCC-MSIF. Section 3 defines the simulation conditions and evaluation metrics for assessing the performance of the proposed method. Section 4 presents the simulation results and discusses comparisons with conventional methods and theoretical limits. Finally, Section 5 states the conclusion of this study.

\section{Principle of Time Delay Estimation}
This section outlines the signal model that forms the basis of TDOA estimation and the principle of time delay estimation based on the cross-correlation method.
First, we summarize the general signal model and the challenges of delay estimation in noisy environments. Next, we describe the basic concept of the generalized cross-correlation (GCC) method \cite{Knapp-1976-GeneralizedCorrelationMethodEstimation,Sim-2016-UnderwaterAcousticSourceLocalization}, which is widely used as a solution, along with representative weighting techniques. Finally, we detail the concept of the GCC-MSIF method proposed in this paper and the derivation process of its specific algorithm. This clarifies the theoretical positioning of the proposed method relative to conventional methods.

\subsection{Generalized Cross-Correlation (GCC) Method}
This subsection describes the fundamental principle of the GCC method and representative conventional methods based on it.
The GCC method improves delay estimation accuracy by applying frequency-domain weighting to the cross-correlation function between observed signals, thereby suppressing the effects of noise propagation. In the following discussion, we distinguish between continuous-time theoretical equations using $\omega$ to represent angular frequency and implementation or discrete processing (such as FFT bins) using $f$ to represent frequency bins. Using this notation clarifies the correspondence between theoretical rigor and the implementation procedure as discrete signal processing.

\subsubsection{Signal Model}
\label{Signal_Model}
Here, we explain the signal model based on conventional studies.
When a signal emitted from a single source $s(t)$ is received by two sensors, the received signals $x_{\rm 1}(t)$ and $x_{\rm 2}(t)$ are defined by the following equations as a model where additive noises $n_{\rm 1}(t)$ and $n_{\rm 2}(t)$ are superimposed, respectively:
\begin{eqnarray}
    x_{\rm 1}(t) &=& s(t) + n_{\rm 1}(t), \label{eq:sig1} \\
    x_{\rm 2}(t) &=& s(t - D) + n_{\rm 2}(t), \label{eq:sig2}
\end{eqnarray}
where $t$ represents time, and $D$ is the signal time difference of arrival between the two sensors. Here, the additive noises $n_{\rm 1}(t)$ and $n_{\rm 2}(t)$ are assumed to be mutually independent.

The objective of TDOA estimation is to estimate the time delay $D$ between the observed received signals $x_{\rm 1}(t)$ and $x_{\rm 2}(t)$. In TDOA estimation, the signal source $s(t)$ is unknown, and the estimated value $\hat{D}$ of the time delay $D$ must be determined from the received signals $x_{\rm 1}(t)$ and $x_{\rm 2}(t)$. Since noise is superimposed on both received signals $x_{\rm 1}(t)$ and $x_{\rm 2}(t)$, this becomes a factor that degrades estimation accuracy.

\subsubsection{Conventional GCC Methods}

Thus, the GCC method has been proposed as a weighting technique to improve TDOA estimation accuracy.
The GCC method estimates the delay time $\hat{D}$ from the peak of the correlation function $R_{\rm gcc}(\tau)$, which is obtained by applying a specific frequency weighting function $\psi_{\rm gcc}(\omega)$ to the cross-spectrum between received signals and performing an inverse Fourier transform. Let $X_{\rm 1}(\omega)$ and $X_{\rm 2}(\omega)$ be the Fourier transforms of $x_{\rm 1}(t)$ and $x_{\rm 2}(t)$, respectively. The generalized cross-correlation function is given by:
\begin{equation}
    R_{\rm gcc}(\tau) = \int_{-\infty}^{\infty} \psi_{\rm gcc}(\omega) G_{12}(\omega) e^{j\omega\tau} d\omega,
    \label{eq:gcc_general}
\end{equation}
where $\omega$ is the angular frequency, $\tau$ is a candidate value for the time delay, and $j$ is the imaginary unit. $G_{12}(\omega)$ is the cross-spectrum between the received signals, defined as:
\begin{equation}
    G_{12}(\omega) = X_{\rm 1}^{*}(\omega) X_{\rm 2}(\omega),
    \label{eq:cross_spectrum}
\end{equation}
where $X_{\rm 1}^{*}(\omega)$ is the complex conjugate of $X_{\rm 1}(\omega)$. The time delay $\tau$ that maximizes this correlation function $R_{\rm gcc}(\tau)$ becomes the estimated value $\hat{D}$ of the delay time $D$.
\begin{equation}
    \hat{D} = \underset{\tau}{\operatorname{arg\,max}} \, R_{\rm gcc}(\tau).
    \label{eq:argmax}
\end{equation}

The performance of the GCC method depends on the selection of the weighting function $\psi_{\rm gcc}(\omega)$ and the spectral averaging process.
Table \ref{tab:gcc_weights} shows representative GCC methods compared in this paper and their weighting functions.
Methods that calculate spectra such as $G_{11}(\omega)$, $G_{12}(\omega)$, and $G_{22}(\omega)$ using averaging are widely used.
This averaging is indispensable in practice for obtaining stable spectral estimates; for instance, methods like Welch's method \cite{Welch-1967-UseFastFourierTransform}, which divides the observed signal into multiple frames and averages them, are often employed.
For methods that utilize the statistical properties of signals and noise, such as the GCC-SCOT and GCC-ML methods described later, this averaging process is a crucial prerequisite.

\begin{table}[htb]
    \centering
    \caption{Weighting functions for representative GCC methods.}
    \label{tab:gcc_weights}
    \begin{tabular}{lc}
        \hline
        Method & Weighting function $\psi(\omega)$ \\
        \hline \hline
        GCC-CC & $1$ \\
        GCC-SCOT & $\cfrac{1}{\sqrt{G_{11}(\omega)G_{22}(\omega)}}$ \\
        GCC-PHAT & $\cfrac{1}{|G_{12}(\omega)|}$ \\
        GCC-ML & $\cfrac{1}{|G_{12}(\omega)|} \cdot \cfrac{|\gamma_{12}(\omega)|^2}{1-|\gamma_{12}(\omega)|^2}$ \\
        \hline
    \end{tabular}
\end{table}

The most basic GCC-CC (cross-correlation) method calculates the correlation directly using the cross-spectrum of the received signals without weighting ($\psi_{\rm cc}(\omega)=1$). Although this method has the advantage of very simple computation, the correlation peak tends to become broad because it is directly affected by the spectral shape of the signal itself and out-of-band noise, resulting in limited estimation accuracy.

The GCC-SCOT (smoothed coherence transform) method \cite{Carter-1973-SmoothedCoherenceTransform} normalizes the cross-spectrum by the square root of the product of the auto-power spectra $G_{11}(\omega)$ and $G_{22}(\omega)$ of the signals. This process has the effect of suppressing the influence when signal spectral energy is concentrated at specific frequencies, thereby sharpening the correlation peak. When applying this method, the aforementioned averaging process is a prerequisite for obtaining stable estimates of the auto-power spectra.

On the other hand, the GCC-PHAT (phase transform) method \cite{Knapp-1976-GeneralizedCorrelationMethodEstimation} effectively extracts only phase information by normalizing with the amplitude component $|G_{12}(\omega)|$ of the cross-spectrum. Through this whitening process, the correlation function approaches an impulse, improving time resolution. Due to this characteristic, the GCC-PHAT method is known to exhibit high robustness, particularly in multipath environments such as indoor settings. The GCC-PHAT method is also referred to as the cross-power spectrum phase (CSP) method.

Finally, the GCC-ML (maximum likelihood) method \cite{Knapp-1976-GeneralizedCorrelationMethodEstimation} provides a theoretically optimal weighting that gives the maximum likelihood estimate of the delay time $D$ under the assumption that the signal and noise follow mutually independent Gaussian processes. This weighting function is defined by the magnitude-squared coherence $\gamma_{12}(\omega)$, and its accurate estimation requires power spectral information of the signal and noise, i.e., stable spectral averaging over multiple frames. While the highest accuracy can be expected if the statistical properties of the signal and noise are accurately grasped, the GCC-ML method is known to have a high computational cost.

\subsection{Proposed Method: GCC with Mean Signal and Inverse Filter (GCC-MSIF)}
This subsection proposes a novel GCC method, termed GCC-MSIF, aimed at improving TDOA estimation accuracy in low SNR environments.
As mentioned in the previous subsection, conventional GCC methods share a common issue: estimation accuracy deteriorates significantly in environments with low signal-to-noise ratios. Therefore, GCC-MSIF aims to construct a method with superior performance in terms of signal-to-noise ratio. In the following, we describe in detail its basic concept, specific formulation, and implementation algorithm.

\subsubsection{Formulation of GCC-MSIF}

In this paper, we propose the GCC-MSIF method aiming to improve accuracy in low SNR environments.
The conventional methods described in the previous subsection share the common problem that estimation accuracy drops remarkably in low SNR environments. The proposed GCC-MSIF is a novel approach that resolves the signal-to-noise ratio issue inherent in conventional GCC methods by introducing the concepts of a mean signal and an inverse filter.
Specifically, it aims to achieve high delay estimation accuracy even in low SNR environments by virtually defining an unknown transmitted signal and performing correlation processing using this virtual transmitted signal.

\begin{figure}[t]
\centering
\includegraphics[width=85mm]{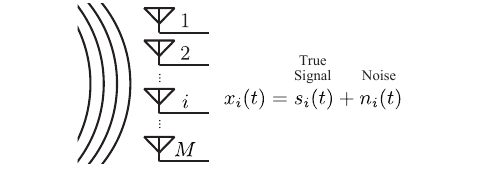}
\caption{Signal model of TDOA estimation. The true signal $s(t)$ propagates from a source and is received by an array of $M$ sensors with individual delays. Independent additive noise $n_i(t)$ is superimposed on each received signal $x_i(t)$.}
\label{ArrayModel}
\end{figure}

The concept of this method lies in utilizing the mean signal of a group of received signals as a virtual transmitted signal.
As shown in the model of the group of received signals in Fig.~\ref{ArrayModel}, the transmitted signal is unknown in TDOA estimation. Therefore, we define the mean signal $x_{\rm mean}(t)$ of the $M$ sensor signals $x_i(t)$ constituting the array by the following equation and consider treating it as a virtual transmitted signal:
\begin{equation}
    x_{\rm mean}(t) = \frac{1}{M} \sum_{i=1}^{M} x_i(t).
\end{equation}
If the noise is uncorrelated, this averaging process is expected to reduce noise components and extract signal components with a high signal-to-noise ratio (SNR).

%%%

However, the mean signal $x_{\rm mean}(t)$ does not serve as a simple substitute (reference signal) for the transmitted signal.
In the mean signal $x_{\rm mean}(t)$, the differences in arrival times to each sensor and the transmission characteristics of each path are averaged. Consequently, using the mean signal simply as a reference signal may result in estimated values that differ from those of the original transmitted signal.
Therefore, to the best of the authors' knowledge, using this mean signal $x_{\rm mean}(t)$ directly as a reference signal has not been practiced.

To address this issue, we introduce a virtual inverse filter $H(f)$ that traces back from the receiving side to the transmitting side.
Consider a filter $H(f)$ that transforms a certain sensor signal $x_i(t)$ into the mean signal $x_{\rm mean}(t)$ (virtual transmitted signal):
\begin{equation}
    H(f) = \frac{X_{\rm mean}(f)}{X_{\rm i}(f)}.
    \label{eq:H_def}
\end{equation}
That is, by multiplying $X_{\rm i}(f)$ by $H(f)$, the output becomes the mean signal $X_{\rm mean}(f)$.
In this paper, we refer to this filter $H(f)$ as an "inverse filter."
While a standard transfer function describes the change from a transmitted signal to a received signal, the filter $H(f)$ is a transfer function from a received signal to a signal intended to be used as a transmitted signal. Since it possesses characteristics opposite to the usual direction, we treat it as a virtual "inverse filter."

\begin{figure}[t]
\centering
\includegraphics[width=85mm]{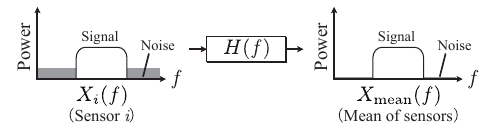}
\caption{Conceptual illustration of the inverse filter in the frequency domain. The inverse filter $H(f)$ transforms the input sensor spectrum $X_i(f)$ into the mean signal spectrum $X_{\rm mean}(f)$. It acts as a band-pass filter that passes signal components while suppressing independent noise components.}
\label{InverseFilterConcept}
\end{figure}

The purpose of the seemingly enigmatic inverse filter $H(f)$ becomes clear when considered per frequency bin.
Figure~\ref{InverseFilterConcept} illustrates the transformation from the received spectrum $X_i(f)$ of sensor $i$ to the mean signal $X_{\rm mean}(f)$ in terms of power spectra.
The behavior of the inverse filter $H(f)$ can be intuitively understood by focusing on each frequency bin. In frequency bins within the signal band, the same signal component is contained in all sensors; thus, the component is preserved even during the averaging process, and the frequency component of $H(f)$ takes a value close to norm 1.
On the other hand, in frequency bins containing noise components that differs for each sensor, the noise power is suppressed by averaging. Since the inverse filter $H(f)$ represents the intensity ratio of the mean signal to the sensor signal, its value approaches norm 0 in frequency bins corresponding to suppressed noise components.
In summary, $H(f)$ functions as a band-pass filter that treats the signal band as a passband and independent noise components as a stopband.
More strictly, since dependent components between sensors (true signal) become the passband and independent components (noise) become the stopband, the inverse filter can be regarded as an "adaptive filter based on dependencies between sensors."

%%%

\begin{figure}[t]
\centering
\includegraphics[width=85mm]{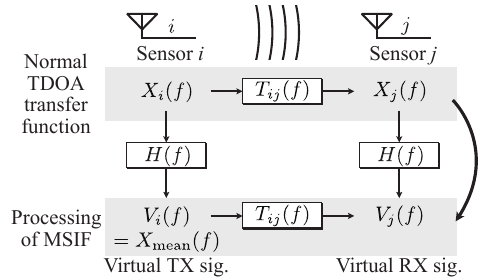}
\caption{Schematic diagram of the MSIF calculation process. The upper path shows the conventional signal propagation model where the transfer function $T_{ij}(f)$ exists between sensor inputs. The lower path illustrates the virtual signal model in MSIF, where the inverse filter $H(f)$ transforms received signals into virtual signals $V_i(f)$ and $V_j(f)$. This process effectively reconstructs a virtual transmission from sensor $i$, enabling correlation with enhanced SNR.}
\label{MSIF_diagram}
\end{figure}

Using this inverse filter $H(f)$, we formulate the correlation processing in the GCC-MSIF method.
Figure~\ref{MSIF_diagram} shows a conceptual diagram of the calculation process of the GCC-MSIF method. Here, the upper tier represents the conventional signal propagation model in TDOA, where $T_{ij}(f)$ denotes the transfer function from sensor $i$ to $j$. The lower tier illustrates the virtual signal propagation model by the GCC-MSIF method.
In the GCC-MSIF method, virtual received signals $V_i(f)$ and $V_j(f)$ are generated by applying the inverse filter $H(f)$ defined in Eq.~(\ref{eq:H_def}) to the received spectra $X_i(f)$ and $X_j(f)$ of each sensor:
\begin{eqnarray}
    V_i(f) &=& X_i(f) H(f) = X_{\rm mean}(f), \label{eq:Vi} \\
    V_j(f) &=& X_j(f) H(f) = X_j(f) \frac{X_{\rm mean}(f)}{X_i(f)}.
    \label{eq:Vj}
\end{eqnarray}
As is clear from Eq.~(\ref{eq:Vi}), the virtual received signal $V_i(f)$ at the reference sensor $i$ becomes the mean signal $X_{\rm mean}(f)$ itself. This can be interpreted as the mean signal being transmitted from the position of sensor $i$.

%%%

By calculating the cross-spectrum between these virtual received signals, we derive the weighting function for the GCC-MSIF method.
The cross-spectrum $G^{\rm msif}_{ij}(f)$ between the virtual received signals is expressed using the conventional cross-spectrum $G_{ij}(f)$ and the inverse filter $H(f)$ as follows:
\begin{eqnarray}
    G^{\rm msif}_{ij}(f) &=& V_i^*(f) V_j(f) \nonumber \\
                         &=& {X}^*_i(f) {H}^*(f) X_j(f) H(f) \nonumber \\
                         &=& |H(f)|^2 X_i^*(f) X_j(f) \nonumber \\
                         &=& |H(f)|^2 G_{ij}(f).
                         \label{eq:G_msif}
\end{eqnarray}
This Eq.~(\ref{eq:G_msif}) indicates that the correlation processing by the GCC-MSIF method is equivalent to applying a weighting of $|H(f)|^2$ to the original cross-spectrum $G_{ij}(f)$, as can be seen from Eq.~(\ref{eq:gcc_general}). That is, the frequency weighting function $\psi_{\rm msif}(\omega)$ of the GCC-MSIF method is defined as the power spectrum of the inverse filter $|H(f)|^2$.

Note that the weighting function $\psi_{\rm msif}(\omega)$ used for implementation can be expressed using observable power spectra.
Substituting Eq.~(\ref{eq:H_def}) into the weighting coefficient $|H(f)|^2$ in Eq.~(\ref{eq:G_msif}) and rewriting it using the definition of the power spectrum $P(f) = |X(f)|^2$, we obtain:
\begin{equation}
    \psi_{\rm msif}(\omega) = |H(\omega)|^2 = \left| \frac{X_{\rm mean}(\omega)}{X_i(\omega)} \right|^2 = \frac{P_{\rm mean}(\omega)}{P_{ii}(\omega)},
    \label{eq:msif_weight_final}
\end{equation}
where $P_{\rm mean}(\omega)$ is the auto-power spectrum of the mean signal, and $P_{ii}(\omega)$ is the auto-power spectrum of sensor $i$.
With this weighting function, the proposed GCC-MSIF method can adaptively select bands with high signal component purity and form sharp correlation peaks, even in low SNR environments.

\subsubsection{Implementation Method}

In this verification, the proposed GCC-MSIF was implemented based on spectral averaging using Welch's method \cite{Welch-1967-UseFastFourierTransform} for practical evaluation and comparison with conventional methods.
Each signal $x_i(t)$ and the mean signal $x_{\rm mean}(t)$ are divided into multiple frames by Welch's method, and the spectral components are stably estimated by frame averaging. This reduces the calculation error of the inverse filter $|H(f)|^2$, leading to improved estimation accuracy.

The weighting function $\psi_{\rm msif}(\omega)$ used for implementation was clipped at 1 to improve estimation stability, based on the theory of $H(f)$.
Ideally, $|H(f)|^2$ becomes 1 in the signal band, but in actual data processing, it may take an excessive value when the denominator is small. To prevent this, after calculating $|H(f)|^2$ for each frequency bin, components where $|H(f)|^2 > 1$ are clipped to 1, thereby reliably avoiding excessive emphasis and instability of estimated values. This policy is an implementation device based on trial and error during the simulation process and is applied only to GCC-MSIF.

\begin{figure}[t]
\centering
\includegraphics[width=155mm]{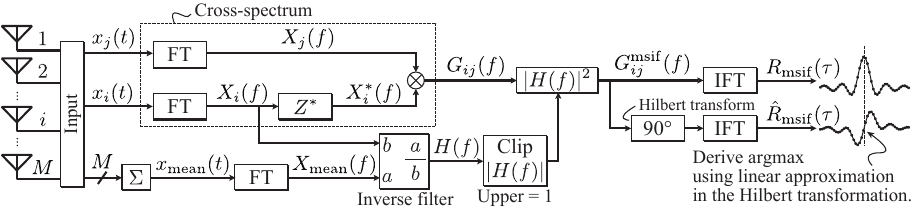}
\caption{
Block diagram of the GCC-MSIF signal processing. Input signals from array sensors are converted to the frequency domain, cross-spectrum and mean signal spectra are calculated, and the MSIF weight $|H(f)|^2$ is applied. The weighted cross-spectrum is transformed to the time domain, and the delay estimate is refined using a Hilbert transform and zero crossing.
}
\label{SignalProcessingDiagram}
\end{figure}

The overall flow of signal processing in GCC-MSIF is shown in Fig.~\ref{SignalProcessingDiagram}.
First, the $M$ sensor input signals are Fourier transformed (FT) after frame division, and the cross-spectrum $G_{ij}(f)$ is calculated for each pair of sensors $i$ and $j$. Furthermore, the mean signal $x_{\rm mean}(t)$ is transformed into the frequency domain to calculate $X_{\rm mean}(f)$. Then, the inverse filter $H(f)=X_{\rm mean}(f)/X_i(f)$ has $|H(f)|$ clipped at 1. The obtained weight $|H(f)|^2$ is multiplied by the cross-spectrum $G_{ij}(f)$ to obtain the MSIF-weighted spectrum $G_{ij}^{\rm msif}(f)$. This is transformed back to the time domain by the inverse Fourier transform (IFT) to obtain the correlation function $R_{\rm msif}(\tau)$.

A technique using the Hilbert transform is employed to achieve high resolution in delay estimation.
The zero-crossing method \cite{Cabot-1981-NoteApplicationHilbertTransform,Grennberg-1994-EstimationSubsampleTimeDelay,Jing-2022-ModifiedAlgorithmBasedQuadratic} was adopted for peak detection of the correlation function $R_{\rm msif}(\tau)$ obtained by GCC-MSIF. The right end of Fig.~\ref{SignalProcessingDiagram} shows an image of the zero-crossing process after the Hilbert transform. This is a method to precisely estimate the peak position $\hat{D}$ with subsample accuracy by performing a Hilbert transform (90-degree phase shift) on the correlation function and extracting the point where the obtained signal crosses $y=0$ (zero crossing) using linear approximation. The zero-crossing method is also applied to the conventional methods used for comparison.

Thus, the main implementation parameters are set uniformly for all methods to ensure a correct comparison with conventional methods.
In the data shown in the subsequent simulations, values such as frame length, window function, averaging conditions, and estimation range, as well as the subsample estimation method, are unified to conduct a fair implementation evaluation. Specific setting values and evaluation conditions are described in detail in Section \ref{SimulationMethodology} (Simulation Methodology).

%%%

\section{Simulation Methodology}
\label{SimulationMethodology}

In this study, we evaluated the performance of the proposed GCC-MSIF method using Monte Carlo simulations to verify its effectiveness.
The evaluation focuses on TDOA estimation on a scale assuming sound source localization for devices such as smart speakers. Specifically, we simulate a "blind" situation where the signal exists only in a specific unknown narrow band amidst broadband white noise across the entire observation band, verifying the robustness of each method against out-of-band noise. This section describes the common simulation conditions and the two evaluation scenarios conducted.

\subsection{Simulation Parameters}

Table \ref{tab:sim_params} shows the main parameters used in the simulation.
A narrowband signal with a center frequency $f_c = 1000~\mathrm{Hz}$ and a bandwidth $f_{\rm BW} = 300~\mathrm{Hz}$ was used as the source, and observation was performed for 0.1 seconds at a sampling frequency $f_s = 48~\mathrm{kHz}$. The noise environment was assumed to be white Gaussian noise with constant amplitude and random phase across the entire frequency band. Regarding the receiving array model, the reception delay at each element was set using uniform random numbers in the range of 0 to 5 samples. However, the true delay difference $D_{ij}$ between the two sensors under evaluation (sensors $i$ and $j$) was fixed at 5 samples (approximately $104~\mu\mathrm{s}$), considering typical delay amounts in applications such as direction of arrival estimation.

For the calculation of evaluation metrics, statistical evaluation was performed with a sufficient number of trials.
Monte Carlo trials were conducted 20000 times for each condition to confirm the error distribution between the estimated delay and the true value. In signal processing, Welch's method was used for transformation to the frequency domain, and a Hamming window (window length 480 samples, overlap rate 75\%) was applied to reduce the variance of spectral estimation.

\begin{table}[htbp]
    \centering
    \caption{Simulation parameters.}
    \label{tab:sim_params}
    \begin{tabular}{lc}
        \hline
        Parameter & Value \\
        \hline \hline
        Sampling frequency $f_s$ & $48~\mathrm{kHz}$ \\
        Signal duration & $0.1~\mathrm{s}$ \\
        Signal center frequency $f_c$ & $1000~\mathrm{Hz}$ \\
        Signal bandwidth $f_{\rm BW}$ & $300~\mathrm{Hz}$ \\
        True delay $D_{ij}$ & $5~\mathrm{samples}$ \\
        Number of Monte Carlo trials & $20000$ \\
        Window function & Hamming \\
        Window length / Overlap & $480$ / $75\%$ \\
        \hline
    \end{tabular}
\end{table}

\subsection{Evaluation Scenarios}

To verify the performance of the proposed method from multiple perspectives, evaluations were conducted in the following two scenarios.
In both scenarios, the in-band SNR ($\mathrm{SNR}_{\rm in}$) was varied from $-14~\mathrm{dB}$ to $+20~\mathrm{dB}$, and the root mean square error (RMSE) was calculated to evaluate tolerance to noise levels. Representative conventional GCC methods, namely GCC-CC, GCC-SCOT, GCC-PHAT, and GCC-ML, were used for comparison, along with the theoretical limit, CRLB. Details on the definition and calculation method of the CRLB used in this evaluation are described in the subsequent Section \ref{CRLB}.

\subsubsection{Scenario 1: Basic Performance Evaluation in White Noise}

In this scenario, we evaluated the estimation accuracy using a sufficient number of array elements ($M=16$) to confirm the basic performance of the proposed GCC-MSIF method.
As mentioned earlier, the objective is to verify whether GCC-MSIF can appropriately emphasize the signal band and improve estimation accuracy in a situation where the signal band is unknown, whereas conventional methods (especially GCC-PHAT and GCC-SCOT) treat information across the entire band equally.

\subsubsection{Scenario 2: Dependency on Number of Array Elements}

In this scenario, we evaluated changes in estimation performance when the number of array elements $M$ was varied as $2, 4, 8, 16$.
Since GCC-MSIF generates a mean signal (virtual transmitted signal) by utilizing multi-channel information in principle, an SNR improvement effect accompanying an increase in the number of elements is expected. By quantitatively evaluating the performance improvement rate for each number of elements, we verify the practical array configuration requirements for the proposed method and its effectiveness with a small number of elements.

\subsection{Theoretical Performance Limit: Cram\'{e}r-Rao Lower Bound (CRLB)}
\label{CRLB}
We introduce the CRLB \cite{Knapp-1976-GeneralizedCorrelationMethodEstimation,Panek-2007-ErrorAnalysisBoundsTime,Chen-2006-TimeDelayEstimationRoom}, which is the theoretical limit of TDOA estimation, as a benchmark for evaluating the performance of the proposed method.
The Cram\'{e}r-Rao lower bound (CRLB) is an index that gives the lower bound of the variance that an unbiased estimator can achieve, indicating the theoretical minimum of delay estimation error in TDOA estimation. The lower bound of the variance $\sigma_{\rm CRLB}^2$ of the delay estimate $\hat{D}$ for an observation time $T$ is given by the following equation \cite{Knapp-1976-GeneralizedCorrelationMethodEstimation} using the magnitude-squared coherence function $\gamma_{12}(f)$:
\begin{equation}
    \sigma_{\rm CRLB}^2 \ge \left[ 2T \int_{0}^{\infty} (2\pi f)^2 \frac{|\gamma_{12}(f)|^2}{1 - |\gamma_{12}(f)|^2} df \right]^{-1}.
    \label{eq:crlb_general}
\end{equation}
This equation indicates that the longer the observation time $T$ and the larger the coherence $\gamma_{12}(f)$ (closer to 1), the smaller the theoretical lower bound of the delay estimation error becomes.

In the simulation of this study, we use the CRLB as a closed-form approximation, assuming SNR and a rectangular spectrum.
First, consider the received signal model defined in the Signal Model (Section \ref{Signal_Model}):
\begin{eqnarray}
    x_{\rm 1}(t) &=& s(t) + n_{\rm 1}(t), \label{eq:sig1_ref} \\
    x_{\rm 2}(t) &=& s(t - D) + n_{\rm 2}(t), \label{eq:sig2_ref}
\end{eqnarray}
where $s(t)$ is the common signal component, and $n_{\rm 1}(t)$ and $n_{\rm 2}(t)$ are mutually independent additive noises. Let $P_s(f)$ be the power spectral density of the signal in the frequency domain. Although the noises $n_1(t)$ and $n_2(t)$ are uncorrelated, for simplicity, we assume their power spectral densities are equal ($P_{n1}(f) = P_{n2}(f) = P_n(f)$). Under these conditions, in the sense of expectation, the auto-spectra and cross-spectrum can be written as:
\(
G_{11}(f) = G_{22}(f) = P_s(f) + P_n(f),\,
G_{12}(f) = P_s(f).
\)
Therefore, the magnitude-squared coherence is expressed as:
\begin{equation}
    |\gamma_{12}(f)|^2 = \frac{|G_{12}(f)|^2}{G_{11}(f) G_{22}(f)}
                       = \frac{P_s(f)^2}{\bigl(P_s(f)+P_n(f)\bigr)^2}
                       = \frac{\eta(f)^2}{\bigl(1+\eta(f)\bigr)^2},
    \label{eq:gamma_snr}
\end{equation}
where $\eta(f) = P_s(f)/P_n(f)$ is the in-band signal-to-noise power ratio (SNR). Furthermore, the ratio appearing in the integrand of the general CRLB equation (\ref{eq:crlb_general}), \(\dfrac{|\gamma_{12}(f)|^2}{1-|\gamma_{12}(f)|^2}\), simplifies using Eq.~(\ref{eq:gamma_snr}) as follows:
\begin{equation}
    \frac{|\gamma_{12}(f)|^2}{1-|\gamma_{12}(f)|^2}
    = \frac{\dfrac{\eta(f)^2}{(1+\eta(f))^2}}
           {1 - \dfrac{\eta(f)^2}{(1+\eta(f))^2}}
    = \frac{\eta(f)^2}{(1+\eta(f))^2 - \eta(f)^2}
    = \frac{\eta(f)^2}{1 + 2\,\eta(f)}.
    \label{eq:coherence_approx_derive}
\end{equation}
In this paper, we assume that $\eta(f)$ is constant within the band where the signal exists and use this value as the in-band SNR.

When a rectangular spectrum is assumed for the simulation, the CRLB can be calculated in a simpler form.
Assume that the signal has a rectangular spectrum with a center frequency $f_c$ and a bandwidth $f_{\rm BW}$, and the observation time is $T$. In this case, the integration in Eq.~(\ref{eq:crlb_general}) is limited to the signal band $[f_c-f_{\rm BW}/2,\; f_c+f_{\rm BW}/2]$, and according to Eq.~(\ref{eq:coherence_approx_derive}), the coherence ratio becomes a constant value within the band:
\(
\dfrac{\eta_{\rm in}^2}{1+2\,\eta_{\rm in}}.
\)
Therefore, the integral term can be analytically obtained as:
\begin{equation}
    \mathrm{I}_{\omega^2} = \int_{f_c - f_{\rm BW}/2}^{f_c + f_{\rm BW}/2} (2\pi f)^2 df
    = (2\pi)^2 \frac{(f_c + f_{\rm BW}/2)^3 - (f_c - f_{\rm BW}/2)^3}{3}.
    \label{eq:integral_w2}
\end{equation}
Substituting Eq.~(\ref{eq:coherence_approx_derive}) and Eq.~(\ref{eq:integral_w2}) into Eq.~(\ref{eq:crlb_general}), the lower bound of the delay estimation error variance, given an in-band SNR of $\eta_{\rm in}$, becomes:
\begin{equation}
    \sigma_{\rm CRLB}^2
    = \frac{1}{\,2T\, \dfrac{\eta_{\rm in}^2}{1+2\,\eta_{\rm in}} \, \mathrm{I}_{\omega^2}\,}.
    \label{eq:crlb_closed}
\end{equation}

In the numerical calculations of this study, the CRLB is evaluated directly using Eq.~(\ref{eq:crlb_closed}).
First, let $\mathrm{SNR}_{\rm in,dB}$ be the in-band SNR (dB value) set in the simulation, and convert this to a linear scale to obtain $\eta_{\rm in}$:
\begin{equation}
    \eta_{\rm in} = 10^{\mathrm{SNR}_{\rm in,dB}/10}.
    \label{eq:snr_linear}
\end{equation}
Next, calculate the coherence term derived in Eq.~(\ref{eq:coherence_approx_derive}):
\begin{equation}
    C_{\rm snr} = \frac{\eta_{\rm in}^2}{1 + 2\,\eta_{\rm in}}.
    \label{eq:coh_scalar}
\end{equation}
Furthermore, calculate the frequency integral value $\mathrm{I}_{\omega^2}$ using Eq.~(\ref{eq:integral_w2}), and finally determine the variance and standard deviation of the CRLB using the following equations:
\begin{eqnarray}
    \sigma_{\rm CRLB}^2 &=& \frac{1}{\,2T \, C_{\rm snr} \, \mathrm{I}_{\omega^2}\,},
    \label{eq:crlb_var_scalar} \\
    \sigma_{\rm CRLB}   &=& \sqrt{\sigma_{\rm CRLB}^2}.
    \label{eq:crlb_std_scalar}
\end{eqnarray}
In the simulations of this paper, the theoretical limits for each in-band SNR are calculated according to the procedure in Eqs. (\ref{eq:snr_linear})--(\ref{eq:crlb_std_scalar}) and used as comparison criteria for the RMSE of GCC-CC, GCC-SCOT, GCC-PHAT, GCC-ML, and the proposed GCC-MSIF.

\section{Results and Discussion}

\subsection{Scenario 1: Basic Performance Evaluation in White Noise}

\begin{figure}[t]
\centering
\includegraphics[width=85mm]{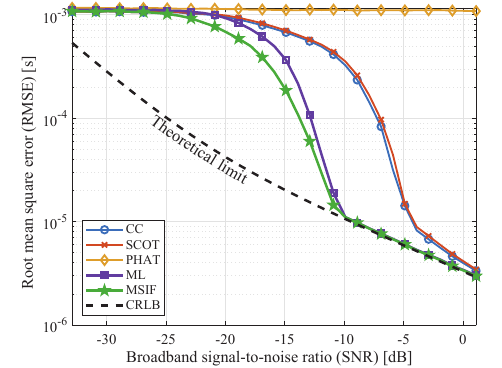}
\caption{RMSE performance of TDOA estimation versus broadband SNR in a blind scenario with broadband white noise. The proposed method (GCC-MSIF) maintains performance close to the CRLB even in low SNR regions where conventional methods fail.}
\label{RMSE_SNR_plot}
\end{figure}

Figure~\ref{RMSE_SNR_plot} shows the RMSE of each GCC method and the proposed GCC-MSIF with respect to the broadband SNR.
Note that the horizontal axis represents the broadband SNR, which is the ratio of signal power to noise power over the entire band, and this differs from the in-band SNR ($\mathrm{SNR}_{\rm in}$) defined in the previous section. Since a narrowband signal is used in this simulation, the broadband SNR has a lower value than $\mathrm{SNR}_{\rm in}$.
The dashed line in the figure represents the theoretical limit (CRLB) derived in Section \ref{CRLB}. In the region where the broadband SNR exceeds $-5~\mathrm{dB}$, GCC-CC, GCC-SCOT, GCC-ML, and the proposed GCC-MSIF all asymptotically approach the CRLB, indicating that sufficient accuracy is obtained regardless of the difference in weighting functions.

The reason why GCC-PHAT did not reach the CRLB even at high SNR is due to the signal conditions of this simulation.
GCC-PHAT applies weighting that uniformly whitens the entire frequency band, which excessively emphasizes noise in bands where the signal does not exist. Under the "blind" condition of this experiment, where a narrowband signal is buried in broadband noise, the influence of out-of-band noise becomes dominant. Consequently, GCC-PHAT failed to perform appropriate estimation even when the in-band SNR was high.

The superiority of the proposed GCC-MSIF became prominent in the low SNR region.
As is clear from Fig.~\ref{RMSE_SNR_plot}, conventional methods such as GCC-CC and GCC-SCOT show a sharp increase in RMSE and a degradation in estimation accuracy when the broadband SNR falls below $-5~\mathrm{dB}$. Note that in the region where estimation completely fails, the RMSE plateaus at approximately $10^{-3}~\mathrm{s}$; this is because the estimated values are constrained within the range of the block processing window length (480 samples). In contrast to this limit, the proposed GCC-MSIF and GCC-ML maintain stable estimation down to lower SNRs. Particularly noteworthy is the position of the "threshold SNR," where the RMSE begins to deteriorate rapidly. In TDOA estimation, there exists a transition region \cite{Weiss-1983-FundamentalLimitationsPassiveTime} where the estimation error increases sharply below a certain SNR. It can be confirmed that GCC-MSIF shifts this threshold by approximately $5~\mathrm{dB}$ to the superior side (lower SNR side), similar to the GCC-ML method.

Furthermore, the proposed GCC-MSIF demonstrated performance equal to or better than GCC-ML, which performs theoretically optimal weighting, even in the low SNR region.
Especially around a broadband SNR of $-15~\mathrm{dB}$, while the RMSE of GCC-ML tends to deteriorate, GCC-MSIF maintains an accuracy closer to the CRLB. This can be interpreted as the process where GCC-MSIF utilizes multi-channel information to generate a "mean signal" and reconstructs it into a virtual transmitted signal using an inverse filter, thereby improving the effective SNR. These results confirm that even under blind conditions, GCC-MSIF data-drivenly emphasizes signal components and effectively suppresses the influence of out-of-band noise.

\subsection{Scenario 2: Dependency on Number of Array Elements}

\begin{figure}[t]
\centering
\includegraphics[width=85mm]{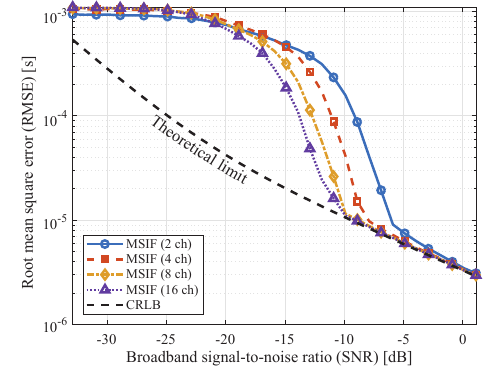}
\caption{RMSE performance of GCC-MSIF with varying number of array elements ($M=2, 4, 8, 16$). Increasing the number of sensors improves the noise robustness, shifting the threshold SNR towards lower values due to the enhanced quality of the estimated mean signal.}
\label{MSIF_ArrayNumberAnalysis}
\end{figure}

It was confirmed that the estimation performance of the proposed GCC-MSIF significantly improves as the number of array elements (channels) increases.
Figure~\ref{MSIF_ArrayNumberAnalysis} shows the RMSE versus broadband SNR when the number of array elements $M$ is varied as $2, 4, 8, 16$. The theoretical limit (CRLB) is shown as a dashed line for comparison.
From the figure, it can be seen that compared to the case of $M=2$ (solid blue line), the RMSE curves shift overall towards the lower SNR side (left side) as the number of elements increases. In particular, focusing on the threshold SNR, which indicates the limit where stable estimation can be performed, an improvement of approximately $10~\mathrm{dB}$ is observed for the case of $M=16$ (purple dotted line) compared to the case of $M=2$.

This performance improvement is due to the noise suppression effect in the "mean signal" generation process of the MSIF method.
In the MSIF method, a virtual transmitted signal (mean signal) is estimated by averaging the signals of all channels, and this is used as a reference signal to design the inverse filter for each channel. As the number of elements $M$ increases, the effect of suppressing random noise by the averaging process increases, improving the SNR of the generated mean signal. Consequently, the estimation accuracy of individual inverse filters is improved, and it is considered that the phase information of the final cross-spectrum is more accurately reconstructed.

It is noteworthy that a clear performance improvement is obtained compared to the two-element case even with only four elements ($M=4$).
This suggests that the proposed method can function effectively not only in large-scale microphone arrays but also in small devices with a limited number of elements, such as smart speakers and robots. From the above results, it has been demonstrated that the proposed method can scalably improve performance according to the number of available channels and is effective in applications requiring robustness, especially in low SNR environments.

\section{Conclusion}

In this paper, we proposed a generalized cross-correlation method using mean signal and inverse filter (GCC-MSIF) as a high-precision TDOA estimation method for passive localization.
The proposed method estimates a virtual transmitted signal (mean signal) from multi-channel received signals and suppresses noise components through inverse filter processing based on it, thereby improving robustness against unknown signal bands and broadband noise. The effectiveness of the proposed method was verified from multiple perspectives through numerical simulations assuming a small array such as a smart speaker.

The main findings obtained in this study are as follows:
\begin{itemize}
    \item \textbf{Superiority in low SNR environments:} In broadband white noise environments, GCC-MSIF maintained stable estimation accuracy even in low SNR regions where conventional methods (GCC-CC, GCC-SCOT, GCC-PHAT) failed. In particular, it was confirmed that the threshold SNR, where estimation error increases rapidly, can be improved to a level equivalent to or better than GCC-ML, which is considered theoretically optimal.
    \item \textbf{Performance improvement with multi-element arrays:} Estimation accuracy was significantly improved as the quality of the mean signal increased with the number of array elements. Clear performance improvement was confirmed even with an array configuration of only four elements compared to the two-element case, demonstrating its effectiveness in small devices.
    \item \textbf{Adaptation to blind environments:} In "blind" situations where the signal frequency band is unknown, while GCC-PHAT suffered performance degradation due to out-of-band noise, the proposed method demonstrated high robustness by data-drivenly suppressing noise bands.
\end{itemize}

On the other hand, when applying this method to larger-scale array systems, consideration of the difference in signal arrival times becomes necessary.
Although this study assumed a small array with narrow element spacing, if the scale of the array increases and the delay difference between received signals becomes large relative to the signal wavelength, simple averaging may cause signal components to cancel each other out. In such cases, coarse time alignment is required as a preliminary step to calculating the mean signal. However, since the purpose here is merely to obtain the outline of the average power spectrum, strict synchronization accuracy is not required, and it is considered that this can be adequately handled with low-complexity processing.

Future work includes verification in more complex real-world environments.
While this paper addressed narrowband signals and white noise environments as a fundamental study, applicability to broadband signals, performance in multipath (reverberant) environments such as those simulated by the image method of Allen and Berkley \cite{Allen-1979-ImageMethodEfficientlySimulating}, and robustness against colored noise and correlated noise remain important verification items for the future. Additionally, detailed computational cost comparison with GCC-ML and real-time implementation using actual hardware are also subjects for future work.

\acknowledgment
This work was supported by NARO Innovation Promotion Program.
NARO stands for National Agriculture and Food Research Organization.

% \appendix
% \section{}
% Use the \verb|\appendix| command if you need an appendix(es). The \verb|\section| command should follow even though there is no title for the appendix (see above in the source of this file).

%%%%%%%%
% \bibliographystyle{apsrev4-1}
% \bibliographystyle{plain} % 仮
\bibliography{mybib}
%%%%%%%%
% \begin{thebibliography}{9}
% \bibitem{apex} The abbreviation for APEX should be ``Appl. Phys. Express'' in the reference list.
% \bibitem{jjap} The abbreviation for JJAP should be ``Jpn. J. Appl. Phys.'' in the reference list.
% \bibitem{instructions} More abbreviations of journal titles are listed in ``Instructions for Preparation of Manuscript'', which is available at our Web site.
% \bibitem{newversion} From jjap3.cls version 2.0 released on April 2011.
% \end{thebibliography}

\end{document}